\documentclass[aps,pra,letterpaper,twocolumn,superscriptaddress,floatfix]{revtex4-1}
\usepackage{graphicx,psfrag,amsmath,amssymb,amsfonts,bbm,latexsym,color,dcolumn,bm,upgreek}
\usepackage{comment}
\usepackage{graphicx}% Include figure files
\usepackage{bm}% bold math
\usepackage{hyperref}% add hypertext capabilities

\begin{document}

\preprint{APS/123-QED}
\title{Thermal Phase Fluctuations in Narrow Superfluid Rings}

\author{Parth Sabharwal}
\affiliation{Department of Physics and Astronomy, Dartmouth College, 6127 Wilder Laboratory, Hanover NH 03766, USA}
\author{Daniel G. Allman}
\noaffiliation
\author{Pradipta Debnath}
\affiliation{Department of Physics and Astronomy, Dartmouth College, 6127 Wilder Laboratory, Hanover NH 03766, USA}
\author{Kevin C. Wright}%
\affiliation{Department of Physics and Astronomy, Dartmouth College, 6127 Wilder Laboratory, Hanover NH 03766, USA}

\begin{abstract}
Using matter-wave interference, we have investigated thermal phase fluctuations in narrow coplanar, concentric rings of ultracold fermionic superfluids. We found that the correlation length decreases with number density, consistent with theoretical expectations. We also observed that increasing the coupling between the rings leads to greater overall coherence in the system. The phase fluctuations increased with a change from periodic to closed boundary conditions as we applied a potential barrier at one point in a ring. These results are relevant for the implementation of proposals to utilize ultracold quantum gases in large and elongated circuit-like geometries, especially those that require deterministic preparation and control of quantized circulation states.
\end{abstract}

\maketitle

\section{Introduction}

Coherence is a fundamental property of superfluids and superconductors that is essential to many of their applications in circuits used for sensing and information processing. However, phase fluctuations can degrade coherence and even destroy long-range order in sufficiently elongated systems, even when the particle-like excitations are three-dimensional and temperatures are well below the critical temperature~\cite{PetrovPhase-FluctuatingPRL2001}. Under these conditions, local macroscopic order can still be maintained over distances shorter than the correlation length, but not globally, and if one or more of the system dimensions exceeds this correlation length the system is described as quasi-condensed~\cite{MoraExtensionPRA2003}. It also becomes challenging to reliably initialize and control supercurrents in circuits, especially since coherence over longer paths can be disrupted by changes in the system's topology and boundary conditions. These challenges are less acute in systems with large phase stiffness such as conventional superconductors~\cite{RaychaudhuriIOP2022}, but can be important in low-carrier density superconductors~\cite{EmeryImportanceNature1995} and in ultracold quantum gases such as those considered in this work.

The first experiments to show the impact of thermal phase fluctuations on elongated three-dimensional (3D) quantum gases were conducted with Bose-Einstein condensates (BECs) of $^{87}$Rb in a prolate harmonic trap with a variable aspect ratio of up to 51. Modulations of the density profile were observed after ballistic expansion, and the power spectrum of these modulations was shown to depend on the aspect ratio of the trapping potential, the atom number, and the temperature~\cite{DettmerObservationPRL2001, Hellweg_Dettmer2001}. Ballistic expansion techniques have been used to characterize phase fluctuations in two-dimensional Bose gases~\cite{ChoiProbing2DPRL2012, SeoScaling2DPRA2014} and to measure the change in coherence in the crossover from three dimensions to one dimension~\cite{ShahProbingPRL2023}. Other experiments have used Bragg interferometry to measure spatial correlation lengths and correlation functions more directly~\cite{HellwegPhaseCorr2003, RichardPRL2003, CacciapuotiSecond-OrderPRA2003, HugbartCoherentEPJD2005}. Phase fluctuations have also been observed using matter-wave interference of two adjacent, independent quantum gases in elongated, linear double-well configurations~\cite{ShinDoubleWellPRL2004, GatiDoubleWellNJP2006, JoMatterWavePRL2007, ScottExploitingPRL2008, ScottQuantifyingPRA2009}. 

This experiment is the first to examine phase fluctuations in an ultracold gas that can be varied between closed (hard-walled) and periodic boundary conditions. The first theoretical analysis of how boundary conditions affect phase fluctuations in a ring geometry was conducted by Mathey \textit{et al.}, who considered the scenario of a weakly-interacting BEC in the hydrodynamic limit where phase fluctuations arise mainly due to excitation of long-wavelength sound modes~\cite{MatheyRingsPhase}. The phase-fluctuating regime was predicted to arise in the limit of low one-dimensional (1D) density, with a simple criterion involving the trap aspect ratio, particle mass, and number of particles.

Although phase variations have previously been measured in ring-shaped quantum gases using matter-wave interference techniques~\cite{EckelPRX2014, CormanQuenchPRL2014, CaiPersistentPRL2022, DelPacePRX2022, AllmanPRA2024}, in those experiments the combination of aspect ratio, 1D density, and temperature were not suitable for studying the phase fluctuating regime, where the correlation length is smaller than half the ring circumference. Our experimental setup allows us to create narrow, concentric superfluid rings of fermionic $^6$Li with aspect ratios $>$10 and uniform 1D densities as low as 20 pairs/$\upmu$m, which is sufficient to realize the phase fluctuating regime. In the hydrodynamic limit of strong interactions, the collective phononic excitations of the superfluid dominate quasiparticle (pair-breaking) excitations, and phase fluctuations should be adequately described by the same Bogoliubov de Gennes (BdG) framework previously applied to weakly-interacting BECs.

\begin{figure}[t]
\includegraphics[width=0.48\textwidth]{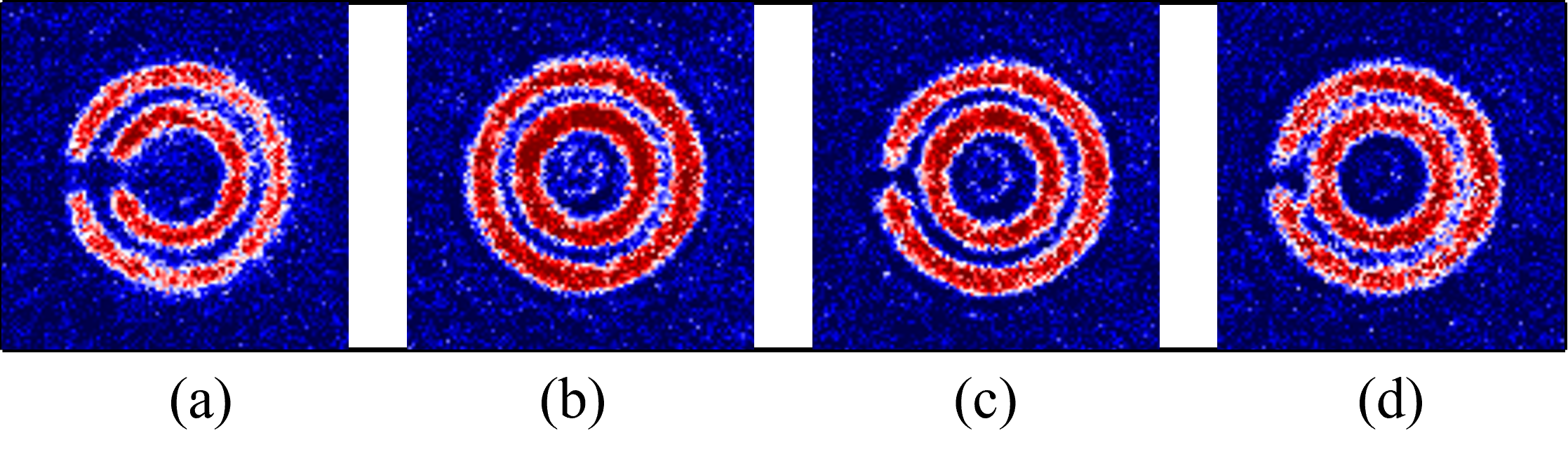}
\caption{Individual absorption images of an equal spin mixture of $^6$Li atoms in narrow concentric rings, showing how the ring geometry and inter-ring coupling is altered for different versions of the experimental procedures. A localized ``cutting'' barrier is applied to both rings during the evaporative cooling process as shown in (a), to ensure the average azimuthal current density is negligible. After cooling, we adiabatically remove the cutting barrier from both rings (b) or just the outer ring (c). These are the `Connected' and `Cut' ring cases described in the text. The smaller inner ring is less affected by fluctuations, and is used as a phase reference for matter-wave interference. We also examined the effects of reducing the strength of the ``inter-ring'' barrier by up to 50\%, which can be seen as a slight increase in the average optical depth between the rings in (d). The field of view is 42~$\upmu$m$\times$42~$\upmu$m.}
\label{ring_images}
\end{figure}

This work focuses on two distinct physical configurations: a pair of smooth and fully connected concentric rings, and a connected inner ring surrounded by an outer ring that is cut at a fixed point by a narrow potential barrier, as shown in Fig.~\ref{ring_images}.
Our motivation for studying configurations with and without localized potential barriers comes from our experiences with using such barriers to initialize narrow superfluid rings in a selected persistent current state. Evaporative cooling through the superfluid phase transition in a smooth, connected ring often leads to the appearance of a persistent current around the ring. This can occur spontaneously via the Kibble-Zurek mechanism~\cite{AllmanPRA2024}, or due to bulk flow generated by changes in the trapping potential during the evaporation process. Our most reliable procedure for initializing two concentric rings each in a zero-current state involves evaporative cooling with a barrier breaking each ring, and reconnecting them only after the temperature is low enough for the superfluid to be robust. The success rate of this initialization procedure is $>99\%$ for sufficiently high final densities and strong interactions, but we also observed that this success rate decreases at lower densities. This work confirms that increased phase fluctuations were responsible for difficulties with initialization, and to determine what other factors might also impact our ability to control the quantized current around the rings.

One essential technical development in this experiment is a robust method for using matter-wave interference to measure the angular variations in the phase difference between adjacent regions of two concentric superfluid rings. The procedure involves using ballistic expansion to generate a suitable matter-wave interference pattern from the two concentric rings, and employing a phase extraction algorithm to infer the phase difference between them. We have validated this experimental measurement procedure using simulated data that combines theoretical predictions of the phase fluctuations with numerical modeling of the matter-wave interference patterns that should appear after ballistic expansion. Sec.~\ref{exptech} contains a detailed description of the experimental setup, the phase extraction algorithm, and a comparison of its effectiveness against the model described in Sec. \ref{theorysim}. In Sec.~\ref{Results}, we describe our results in detail and compare them with theoretical predictions, and discuss the limitations of our study.

\section{Experimental Techniques/Methods} \label{exptech}

The goal of the experiment is to study phase fluctuations of a $^6$Li superfluid confined in a double ring configuration, as we vary the pair densities and inter-ring coupling. We prepare the $^6$Li atoms in the trap in an equal spin mixture of the lowest two hyperfine states, which have a broad Feshbach resonance occurring at $83.4$~mT. This work is focused on the phase fluctuating properties of the fermionic pair superfluid in the unitary limit where the superfluid critical temperature is relatively high, and it is reasonable to treat the system in the hydrodynamic limit.

Core technical details of the cooling, trapping and state preparation procedure are described in detail in Refs.~\cite{CaiPersistentPRL2022, AllmanPRA2023, AllmanPRA2024}, but the final part involves evaporative cooling of the $^6$Li atoms in a dynamically reconfigurable optical dipole trap generated by two red-detuned laser fields. The main contribution to the trap potential is from a sheet-like trapping beam with a wavelength of 1064~nm, propagating in the horizontal direction. The broad, oblate harmonic potential near the focus of this beam is modified by a second beam propagating upward, perpendicular to the sheet beam, with a wavelength of 780~nm and a typical power of 50-100~mW. The spatial intensity profile of this beam is shaped by a digital micromirror device (DMD, Texas Instruments DLP Lightcrafter 6500) with a feature resolution limit of 1.5~$\upmu$m, after which 1.5-3~mW of power reaches the atoms. In this experiment we use the DMD to generate the concentric-ring patterns, with the resultant density profiles shown schematically in Fig.~\ref{ring_images},  where the radii of the inner and outer rings are 7.5~$\upmu$m and 12.5~$\upmu$m, respectively, and their Gaussian widths are 3~$\upmu$m each. In the cases where we cut the rings with a localized azimuthal barrier, the width of the cutting barrier is 3~$\upmu$m. The final vertical trapping frequency of our sheet-like trap after evaporative cooling is 1.4(1)~kHz, and the oscillation frequency for radial motion about the minima of the rings is 5.1(2)~kHz. The peak Fermi energy~\cite{FermiEnergyNote} lies within the range $h\times8.7-13.4$~kHz ($h$ is Planck’s constant) for the atom numbers observed in this experiment.

Previous experiments have measured spatial variations of the superfluid phase in a ring configuration using interferometric techniques with bosonic~\cite{EckelPRX2014, CormanQuenchPRL2014} and fermionic~\cite{DelPacePRX2022} condensates in a ``bulls-eye" configuration with a disk surrounded by a ring. Instead, we use a double ring configuration, similar to that employed in~\cite{AidelsburgerPRL2017, AllmanPRA2024}, replacing the disk with an inner ring. The goal is to maximize the sensitivity and accuracy of the interferometric measurement by achieving the best fringe contrast in the shortest possible expansion time after release from the trap. The total atom number in the inner ring matches that of the outer ring more closely than a disc of a similar radius and density, which has been empirically determined to provide much better fringe contrast. This is especially true for fringes formed in the interior region of the disk, where the presence of a large background of atoms reduces the fringe contrast greatly. Using a smaller disk to address this atom number mismatch causes a large increase in the time-of-flight (ToF) before the fringes begin to form. A larger ToF increases the degree of self-interference of neighboring sections of the ring by the end of the expansion, effectively blurring the azimuthal phase information. An inner ring may provide a less uniform phase reference than a disk, but we account for these phase fluctuations in our analysis, as described in Section~\ref{theorysim}. The dimension of the outer ring is itself set by the desired aspect ratio, and the estimated correlation length in our experiment.

We now discuss how we vary the density and inter-ring coupling, which is essential for obtaining the results presented in later sections. We vary the final number of atoms in the trap between $4$-$12\times10^3$ by changing the number of atoms initially loaded into the dipole trap before forced evaporative cooling, which is achieved by lowering the sheet beam power from $3.0$~W to $42$~mW over 2 seconds. We vary the amount of inter-ring coupling, by weakening what we will refer to here as the ``inter-ring barrier". This is achieved by programming the DMD to alter the amount of light in the inter-ring region during the evaporation stage and thereafter, which in turn varies the number of non-condensed atoms in that region and effectively varies the amount of coupling between the two ensembles. We have validated that the column density in the inter-ring region grows linearly with the weakening of the inter-ring barrier.  We note that the possible existence of pre-formed pairs in the inter-ring region could lead to non-trivial modifications of the coupling dynamics. A barrier strength of 100$\%$ corresponds to no red-detuned light in the inter-ring region from the DMD-controlled beam, which is reduced to 50$\%$ in increments of 10$\%$, by increasing the amount of red-detuned light. After evaporation and a hold time of 100~ms, we reconnect one or both rings by adiabatically removing the cutting barrier(s) over another 100~ms, by dynamically updating the pattern on the DMD, resulting in a configuration with a `Cut' or `Connected' outer ring, respectively.

After preparation of the superfluid in the concentric trap potentials described above, we measure the superfluid phase difference profile via matter-wave interference (See Sec.~\ref{phasemeasurement}). To do this, we shut off the optical trapping beams and rapidly ramp the magnetic field down to $61$~mT, which converts most of the weakly-bound pairs at resonance to tightly bound molecules before they are allowed to expand~\cite{RegalPRL2004}. This ensures that the pairs do not break and their coherence is preserved after they are released, which is essential for the initial phase variations to lead to observable density variations in the pattern that appears when the concentric expanding rings begin to overlap and interfere with each other. All imaging of the density of the molecules after expansion is performed by resonant absorption imaging of atoms in the $|1\rangle$ state on a closed transition at $61$~mT.

\subsection{Phase Measurement Procedure}\label{phasemeasurement}

As demonstrated by Eckel \textit{et al.}~\cite{EckelPRX2014} and Corman \textit{et al.}~\cite{CormanQuenchPRL2014}, when two sufficiently independent ring condensates expand and overlap, observable interference fringes can form under certain conditions. These fringes contain information about the in-trap state of each superfluid, with the assumption that initial local phase variations map linearly onto observable deviations in the fringe positions. For two non-circulating, concentric ring condensates in equilibrium, this resultant pattern should look like a series of concentric closed loops, with the polar profile $r(\theta)$ of each loop dependent on the local angular phase difference between the condensates before they were released from the trap. If there is a non-zero winding number difference $\Delta w$ between the two condensates, spirals are seen in the resultant interferogram, with the number of arms equal to $|\Delta w|$. The deviations of the fringes about the expected radial positions indicate fluctuations in the ``relative" phase about the overall winding value. In-trap variations in the local chemical potential also give rise to deviations in fringe positions, although this effect is typically stable shot-to-shot, and small compared to the effect of phase fluctuations on the fringe positions. We account for these deviations by subtracting the offset apparent in the average phase extracted for all the experimental runs considered in our experiment. Our entire analysis is predicated on accurate extraction of the azimuthal variation in phase from an interferogram and relating it back to the phase around the experiment ring.

\begin{figure}[t]
\includegraphics[width=0.48\textwidth]{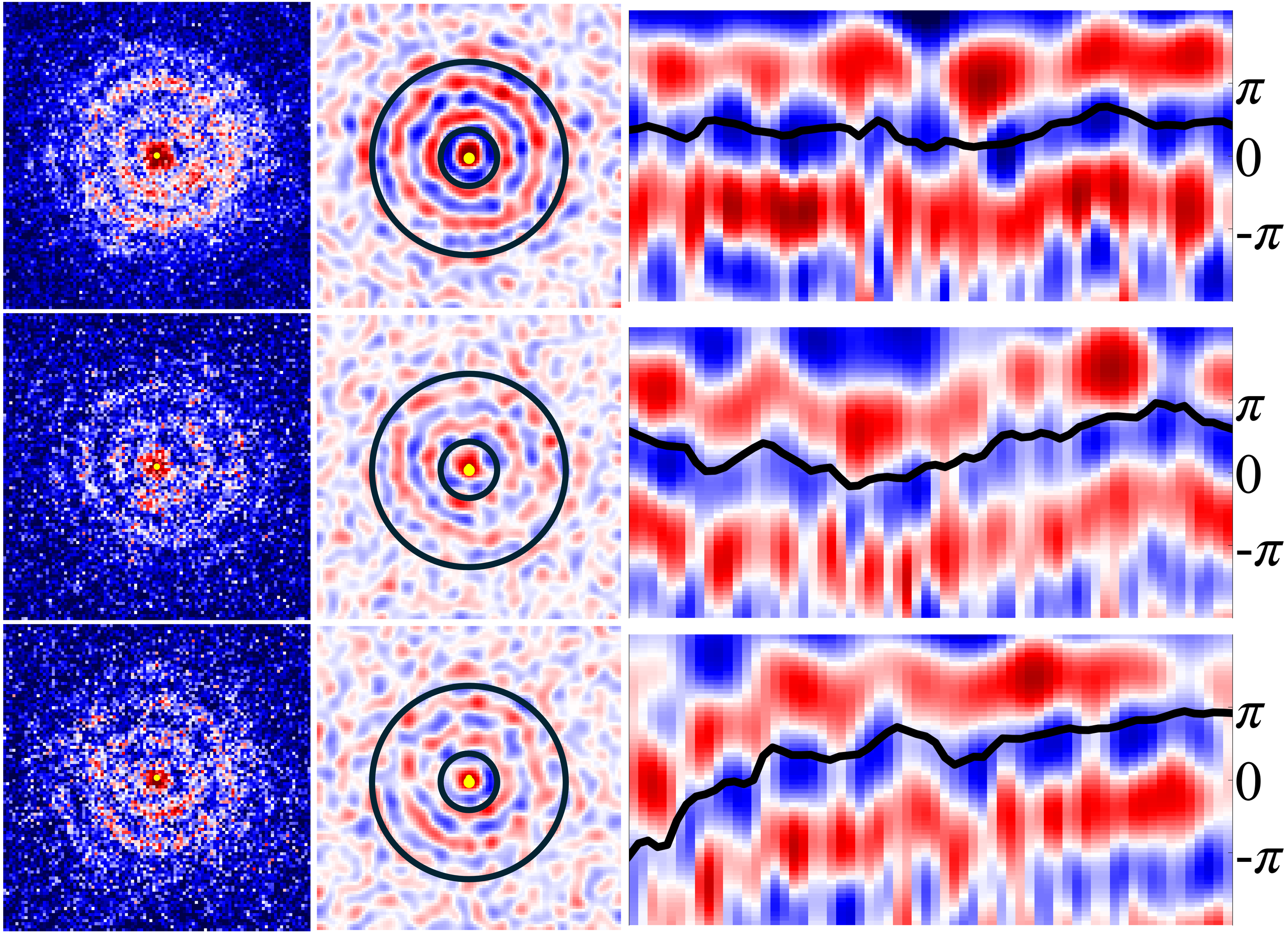}
\caption{Interferograms (left) with a Gaussian background subtracted off (center), accompanied by their polar projection (right) over a range of radii (enclosed by black circles), about the center (yellow dot). This polar image is used for extracting relative phase (black line), which is scaled to match 2$\pi$ to the typical fringe width (right axis), with an arbitrary offset applied. The interferograms shown are from a pair of connected rings with (top), 4.2~$\times 10^3$ (middle) 1.9~$\times 10^3$ and (bottom) 2.1~$\times 10^3$ pairs, with the last case showing an overall relative winding number of 1. The color scale for the interferograms (left) has been normalized to the respective peaks. The field of view is 42~$\upmu$m$\times$42~$\upmu$m.}
\label{Extracted Phase}
\end{figure}

We automate extraction of the \textit{in-situ} relative phase from the interferograms by using a Fourier technique. The first step is the accurate determination of the centre of symmetry, which is determined by locating the central interference peak in the interferogram. To improve contrast and facilitate detection of the interference fringes, we perform a Gaussian background subtraction with a filter radius of 1.7~$\upmu$m, which is on the order of the fringe width 2.1~$\upmu$m, but smaller than the fringe spacing 4.4~$\upmu$m. This radius was chosen empirically to remove the lowest spatial frequency components from the Fourier spectrum while preserving enough of the desired signal at the spatial frequency of the fringes to ensure a robust measurement. 

An annular region of this background subtracted image containing the fringes and excluding the central peak is then mapped from the $x$-$y$ plane to the $\theta$-$r$ plane, effectively ``unwrapping" the fringes present in the original interferogram, as can be seen in Fig.~\ref{Extracted Phase}. The radial extent of the annular region is chosen to be 3.8~$\upmu$m to 13.0~$\upmu$m from the center, and is twice the typical radial fringe spacing. The value of this spacing depends primarily on the chosen ToF and the radii of the inner and outer rings, and is otherwise very stable against changes in experimental conditions. At the optimal ToF, any radial cut of this ``unwrapped" image is close to sinusoidal, and a Fourier transform of this 2D image along the radial axis shows a spectral peak at the second harmonic. For each angular position $\theta$, we extract the phase argument of this Fourier component, which we treat as an estimate of the \textit{in-situ} relative azimuthal phase of the rings, plus a constant phase related to the overall radial offset. We subtract this average offset under the assumption that the average relative azimuthal phase is zero. This routine effectively generates a single composite fringe from the interferogram that can be sampled on the interval $\theta\in[0,2\pi)$. Fig.~\ref{Extracted Phase} shows the extracted phase for a few experimental runs starting with two connected rings for various atom numbers and relative winding numbers ($\Delta w$). It can be seen that the algorithm works well across various atom numbers, and can be effectively used to extract phase and winding number information for large sets of data.

\begin{figure}[t]
\begin{center}
\includegraphics[width=0.44\textwidth]{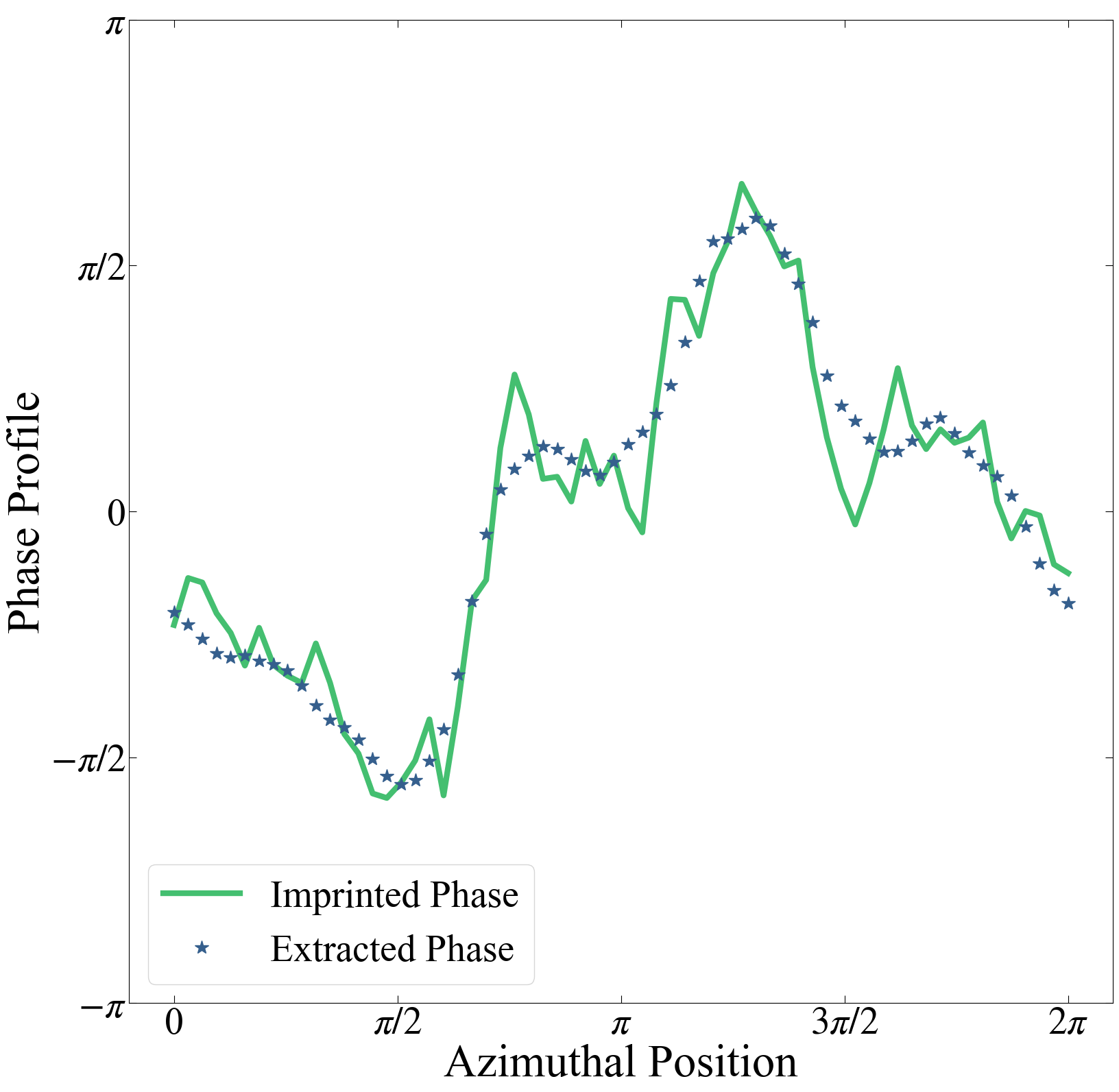}
\caption{Comparison between the phase imprinted on a simulated absorption image (solid line), and extracted using the method described above (stars). Both plots represent the phase difference between the inner and outer ring at each azimuthal position. The correlation length assumed to simulate this instance of imprinted phase is 0.3 times the circumference of the outer ring.}
\label{extractionvalidation}
\end{center}
\end{figure}

\begin{figure*}[t]
\centering
\includegraphics[width=0.95\textwidth]{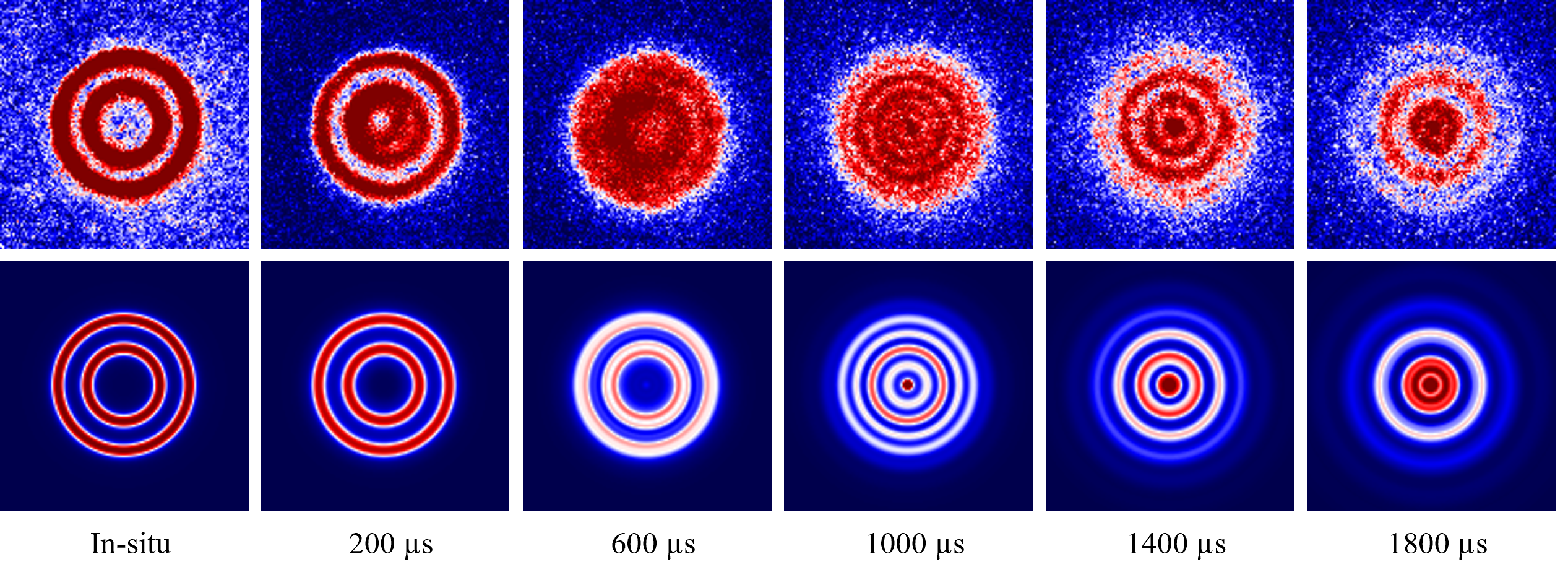}
\caption{Comparison between experimental observations of ballistic expansion of atoms after release from a double ring (top row) against numerical simulations (bottom row), for an overall winding number difference $\Delta w=0$. Each image is an average of 5 absorption images. The radii of the inner and outer ring in these images is 7.5 and 13.5~$\upmu$m, with Gaussian half-widths of 1.5~$\upmu$m. The most optimal fringes for phase extraction are seen here at 1400~$\upmu$s. For the rest of the experimental data, a smaller radius for outer ring (12.5~$\upmu$m) is used, which reveals similar optimal fringes at 1100~$\upmu$s. The field of view is 50~$\upmu$m$\times$50~$\upmu$m.}
\label{evolutionsimulation}
\end{figure*}

We validated the phase extraction procedure outlined above by extracting the phase from interferograms simulated by ballistically propagating two independent superfluids initially confined to two concentric rings and imprinted with a known, controllable phase difference. We find that the assumption of ballistic expansion is sufficient to describe the main features of the interference pattern throughout the relevant times of flight, as can be seen in Fig.~\ref{evolutionsimulation}. A more thorough treatment would account for interactions to capture additional complexities in the expansion dynamics, but this was not essential for validating our phase extraction routine. We perform simulations on a spatial grid of size 50~$\upmu$m$\times$50~$\upmu$m, with a grid spacing of 0.42~$\upmu$m, chosen because a single camera pixel in our EMCCD (Andor Ixon Ultra) represents an extent of 0.42~$\upmu$m in the atomic plane. The dimensions of the rings are identical to those chosen in our experiment. Azimuthal phase variations for each ring are simulated using the model described in the Sec.~\ref{theorysim} for various correlation lengths, which are then imprinted on the superfluid wavefunctions. After propagating the initial wavefunctions with imprinted phases using standard Fourier techniques, the resulting density profile is convolved with an Airy disk of radius 1.5~$\upmu$m to account for the finite resolution of our imaging objective. 
As exemplified in Fig.~\ref{extractionvalidation}, the algorithm substantially recovers the imprinted phase, with better accuracy for the long wavelength components. Crucially, phase variations on length scales comparable to the correlation length are preserved and effectively extracted. The root-mean-square deviations from the imprinted phase are typically less than 0.3 radians for various correlation lengths.

The evolution of the matter-wave interference pattern after the atoms are released from the trap can be quite complex, as can be seen in the images shown in Fig.~\ref{evolutionsimulation}. There are many factors involved in determining the optimum expansion time for obtaining information about the superfluid phase from the concentric rings of fringes appearing in the patterns. This optimum time is generally expected to occur when the two rings have expanded enough to interfere substantially over a large region, but not so much that the inner ring has expanded through the center. For a ring width $d=3$~$\upmu$m, a radial separation of $\Delta r=5$~$\upmu$m, and a pair mass $m_p$ we can estimate that the optimum ToF will be around  $t_\text{ToF}=m_pd\Delta r/\hbar \approx$ 1.4~ms, which is close to our chosen ToF of 1.1~ms. The characteristic radial fringe spacing for the interference of molecular matter-wave components with relative velocity $\delta v=\Delta r/t_\text{ToF}$ is set by the de Broglie wavelength $\Lambda=h/m_p\delta v\approx7.3$~$\upmu$m. For the conditions of our experiment, the fringe spacing is around 5~$\upmu$m, which is easily resolvable using absorption imaging. Finally, for the sake of reliable automated phase-extraction, it is important to employ a ToF such that one observes roughly equal-spaced interference fringes (at fixed polar angles) such that radial slices of the density profile have a single prominent sinusoidal component. For the conditions used in the experiment, we found this to always be true.

\section {Phase fluctuations in a superfluid ring} \label{theorysim}

In this section, we present a simple model for phase fluctuations arising from thermal excitations of Bogoliubov-Anderson (BA) sound modes for a superfluid in a ring geometry. Our main goal is to obtain expressions for the auto-correlation function that explicitly account for the effects of periodic or hard-wall boundary conditions in our ring-shaped channel. We adopt the same Bogoliubov-de Gennes (BdG) approach applied to a weakly-interacting BEC in Ref~\cite{MatheyRingsPhase}, under the assumptions that the hydrodynamic approximation is still valid for the elongated unitary Fermi gas in our experiment, the superfluid fraction is not small, and the temperature is high enough compared to the phonon energy that quantum fluctuations are negligible. We also assume that the interaction energy is large enough that density fluctuations are of negligible importance at distances much larger than the healing length. A more accurate and comprehensive treatment of fluctuations is possible using the framework of a low-temperature effective field theory for fermions~\cite{KliminSpringer2015}, but the simpler BdG approach should suffice for present purposes. We use it to make predictions about the correlation function, and to generate simulated experimental data for validation of our overall phase extraction and analysis procedures.

Under our current experimental conditions, it is clear that the correlation length is always large compared to the width of the ring-shaped channels, and it is reasonable to assume that excitation of transverse sound modes is negligible. In this case a one-dimensional description is valid, and the long-wavelength BA sound modes along the channel may be treated as independent Gaussian-distributed variables with a linear phononic dispersion. For a connected ring-shaped channel of radius $R$, we may expand the phase $\phi_\text{con}(\theta)$ of the pair order parameter in a Fourier series in the angular coordinate $\theta$
\begin{align}\label{ring mode expansion}
    \phi_\text{con}(\theta) &= \sqrt{\frac{R}{ l_\phi }} \sum_{m\neq 0} \frac{\xi_m }{|m|}\frac{e^{i m\theta}}{\sqrt{2\pi}} + c.c.,
\end{align}
where the $\xi_m$ are complex-valued Gaussian normal and independently distributed variables, which satisfy $\langle\xi_m\rangle=0$ and $\langle\xi_m^*\xi_{m'}\rangle=\delta_{mm'}$ with the brackets denoting an ensemble average. The correlation length $l_\phi$ describes the length scale over which the correlation function \mbox{$g_1(\theta)\propto\langle\exp\{i[\phi(\theta)-\phi(0)]\}\rangle$} decays. We explicitly compute the closely-related auto-correlation function later in this section and show that $l_\phi$ indeed represents a correlation length.

This formalism can be adapted to the scenario of the cut ring, keeping in mind the constraint from the boundary condition that $\boldsymbol{\nabla} \phi$ must vanish at the break in the ring. Physically, this means that there is no superflow into or out of the break in the ring, which is required by mass conservation. Therefore, the phonon spatial mode functions in this case are cosinusoidal, and the phase function for the cut ring can be expressed as
\begin{align} \label{cut mode expansion}
    \phi_\text{cut}(\theta) &= \sqrt{\frac{R}{l_\phi }} \sum_{m > 0} \frac{\xi_m }{m/2} \frac{\cos(m\theta/2)}{\sqrt{\pi}}+ c.c.,
\end{align}
where $\theta \in [0,2\pi)$ and there is a break in the ring with zero angular width at $\theta=0$.

In a narrow torus with harmonic transverse confinement, and under the condition that the correlation length is larger than the transverse widths of the toroidal channel, the thermal correlation length of a weakly interacting molecular BEC is given by~\cite{MatheyRingsPhase}
\begin{equation}\label{correlationlenrelation}
l_\phi=\frac{2 \hbar^2}{m_p k_B T}\frac{N_0}{L},\\
\end{equation}
where $T$ is the temperature, $L$ is the channel length, $m_p$ is the molecule pair mass and $N_0$ is the condensed molecule number. Notably, in this approximation the correlation length is expected to depend on temperature and the 1D density $n_0\equiv N_0/L$, but not explicitly on the interaction strength.

Assuming density fluctuations are negligible, we can write general expressions for the angular auto-correlation function in terms of a generic Gaussian-random phase $\phi(\theta)$, defined for a system with either periodic or closed (hard-walled) boundary conditions. For the periodic case, the angles $\theta$ and $\theta+2\pi$ represent physically indistinguishable points, and we define the periodic normalized angular auto-correlation function
\begin{flalign}
    g_{\text{periodic}}[\phi](\theta') & \equiv\frac{\int_{0}^{2\pi} \langle e^{i[\phi(\theta)- \phi(\theta + \theta')]}\rangle d\theta}{\int_{0}^{2\pi} d\theta}\label{autocor_equation1}\\ 
    & = \frac{1}{2\pi} \int_{0}^{2\pi} e^{-\langle[\delta\phi(\theta, \theta')]^2 \rangle/2} d\theta,\label{autocor_equation_periodic}
\end{flalign}
where $\delta\phi(\theta, \theta')\equiv\phi(\theta)-\phi(\theta+\theta')$ and the form in Eq.~\eqref{autocor_equation_periodic} is obtained using Wick's theorem. For the case of closed boundary conditions, the phase is only defined over the extent of the channel, and so we define the closed normalized angular auto-correlation function
\begin{flalign}
    g_{\text{closed}}[\phi](\theta') & \equiv\frac{\int_{0}^{2\pi - \theta'} \langle e^{i[\phi(\theta)- \phi(\theta + \theta')]}\rangle d\theta}{\int_{0}^{2\pi - \theta'} d\theta}\label{autocor_equation2}\\ 
    & = \frac{1}{2\pi - \theta'} \int_{0}^{2\pi-\theta'} e^{-\langle[\delta\phi(\theta, \theta')]^2 \rangle/2} d\theta.\label{autocor_equation_linear}
\end{flalign}
With this definition $\delta\phi(\theta,
\theta')$ is unique and well-defined on the interval $\theta\in[0,2\pi-\theta']$ for $\theta'\in[0,2\pi)$. 

Explicit expressions for auto-correlation functions can be calculated for the connected and cut ring (with a barrier of negligible angular extent) using Eqs.~\eqref{ring mode expansion} and~\eqref{cut mode expansion} for the connected and cut cases, respectively:
\begin{flalign}
    g_\text{con}(\theta';l_\phi,R)&\equiv g_\text{periodic}[\phi_\text{con}](\theta')= e^{-|x|(1-|x|)/\Tilde{l}_\phi}\\
    g_\text{cut}(\theta';l_\phi,R)&\equiv g_\text{closed}[\phi_\text{cut}](\theta')= e^{-|x|/\Tilde{l}_\phi},
\end{flalign}
where $x\equiv \theta'/2\pi$ and $\Tilde{l}_\phi\equiv l_{\phi}/2\pi$\textit{R} are the scaled angular offset and scaled correlation length respectively, and the dependence on the correlation length has been made explicit. While the final expression for the connected ring is the same for both Eqs.~\eqref{autocor_equation_periodic} and  \eqref{autocor_equation_linear} due to translational invariance, the Eq.~\eqref{autocor_equation_linear} is used for the cut ring case.

Having calculated the correlation functions for a single ring, we now use these expressions to determine the auto-correlation function of the experimentally-measurable relative phase \mbox{$\phi_\text{rel}(\theta)\equiv\phi_1(\theta)-\phi_2(\theta)$} between two concentric rings (labelled 1 and 2) of radii $R_1$ and $R_2$ that may either be cut or connected. We take, without loss of generality, \mbox{$R_1<R_2$} and refer to ring 1 as the inner ``reference" ring and ring 2 as the outer ``experiment" ring. Under the assumption of uncorrelated inter-ring phases, i.e. \mbox{$\langle\phi_1(\theta_1)\phi_2(\theta_2)\rangle=0$}, for two rings of the same type, this function takes a form similar to the single-ring expressions, with an apparent scaled correlation length satisfying $\Tilde{l}_{\phi,\text{rel}}^{-1}=\Tilde{l}_{\phi,1}^{-1}+\Tilde{l}_{\phi,2}^{-1}$. For the case of a double ring consisting of a connected reference ring and cut experiment ring, we compute the auto-correlation function using Eq.~\eqref{autocor_equation_linear} since the relative phase is ill-defined at the cut. Due to the independence of the rings, it is easy to show that the auto-correlation function is simply the product of the single-ring auto-correlation functions:
\begin{align}
    g_\text{rel}(\theta')&=g_{\text{con}}(\theta',l_{\phi,1},R_1)g_{\text{cut}}(\theta',l_{\phi,2},R_2).
\end{align}
Accounting for a finite barrier of angular extent $2\pi f$ located at $\theta=0$, the expression is modified as follows:
\begin{align} \label{autocorexplicitexpressionfinal}
    \begin{split}
        g_\text{rel}(\theta')&= \exp\left\{-\frac{|x|(1-|x|)}{\Tilde{l}_{\phi,1}}-\frac{|x|}{(1-f)\Tilde{l}_{\phi,2}}\right\}.
        \end{split}
\end{align}
In our experiment, $2\pi f=0.24$ radians. Furthermore, the domain of the auto-correlation function is restricted to the angular extent of the cut ring, i.e., $\theta'\in[0,2\pi(1-f))$.

For all of the data taken in this experiment, the average 1D densities of each sub-ring are not exactly equal, and this must be included in the model to make a quantitative comparison against the data. In the case where the reference ring is connected and the experiment ring is cut, the 1D density of the reference ring, $n_{0,1}$, is 1.2 times that of the experiment ring, $n_{0,2}$. 
Eq.~\eqref{correlationlenrelation} then implies that the correlation length of the inner ring is a factor $\eta=1.2$ times that of the outer ring, with the assumption that the temperature is uniform throughout the system. With this information, we express our final fitting function as
\begin{align} \label{autocorexplicitexpressionfinal2}
    \begin{split}
        g_\text{rel}(\theta')&= \exp\left\{-\frac{1}{\Tilde{l}_{\phi,2}}\left(\frac{|x|(1-|x|)R_1}{\eta R_2}+\frac{|x|}{1-f}\right)\right\}. 
        \end{split}
\end{align}
We can further substitute the measured scaled correlation lengths into the expression given in Eq. \ref{correlationlenrelation}, to obtain a temperature estimate $T$, where $N_\text{tot}$ is the total number of pairs observed in the interferogram:
\begin{align}\label{temp_equation}
    T&=\frac{2 \hbar^2}{m_p k_B (2\pi R_2 \Tilde{l}_{\phi,2})}\frac{N_\text{tot}}{(2\pi \eta R_1+2\pi R_2)}.
\end{align}
In the above, we have implicitly assumed that $N_\text{tot}$ is the same as the number of pairs initially in the rings. We have confirmed this to be true to within approximately $10\%$ in our experiments by comparing the average \textit{in-situ} pair numbers to those observed in the interferograms. This assumption is important to be able to associate a definite pair number to a particular realization of an observed phase profile without precise knowledge of the \textit{in-situ} state. Only the section of the interferograms containing the fringes, which is a circle of radius 21~$\upmu$m, is considered for these quantitative comparisons.

\section{Results} \label{Results}

\begin{figure}[!t]
\begin{center}
\includegraphics[width=\columnwidth]{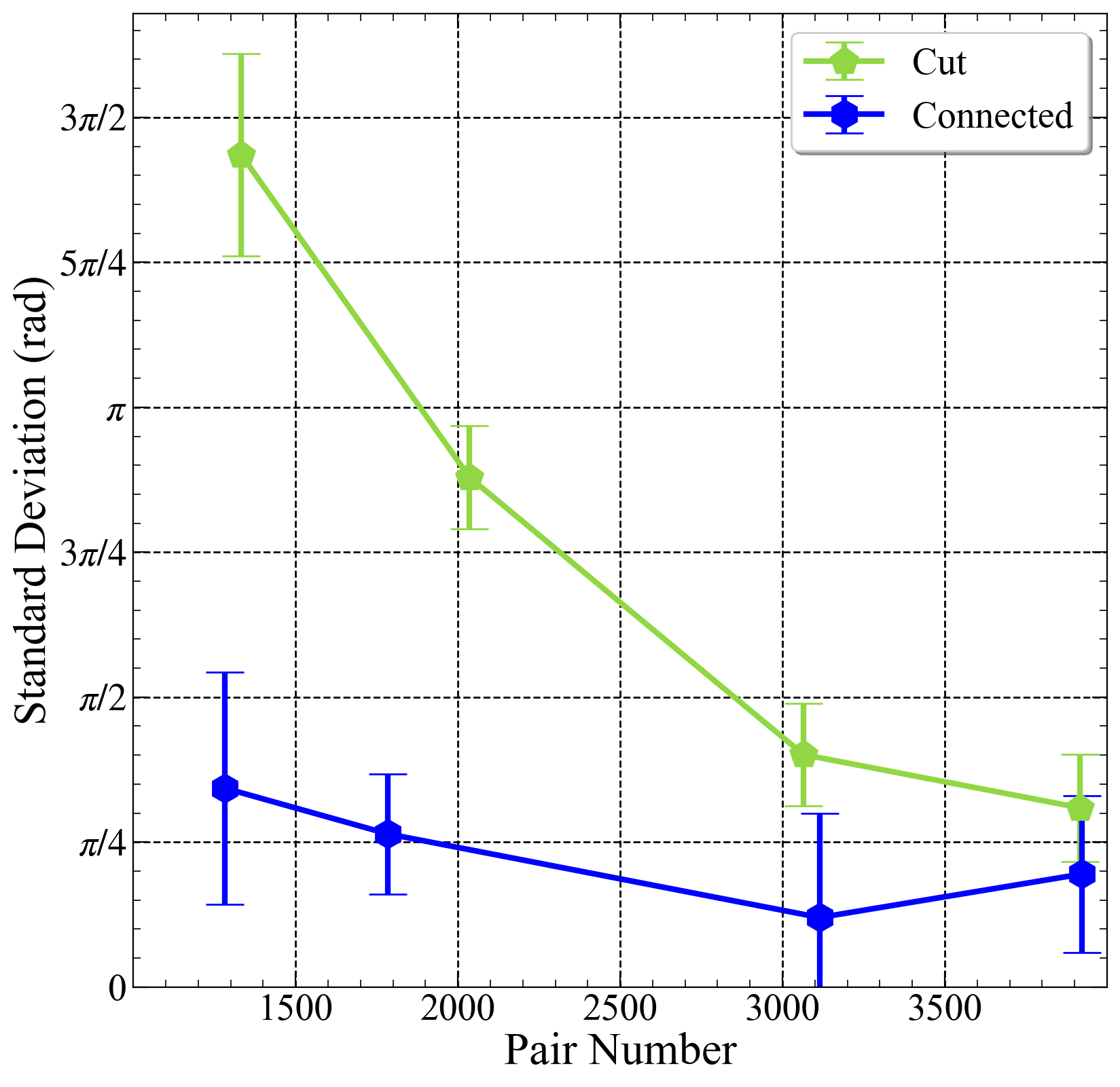}
\caption{Standard deviation of phase difference across an angular region of total width 0.78 radians for an outer ring with (green, `Cut') or without (blue, `Connected'), the cutting barrier. The inter-ring barrier is set to 100$\%$ for all these experimental runs. We observe a dramatic increase in the variance for pair numbers below ~2500. Runs with a non-zero overall winding number for the connected ring have been excluded here. The error bars represent the standard error of the sample standard deviation shown here~\cite{stddevnote}.}
\label{fig:flucvsloadvariance}
\end{center}
\end{figure}

The first investigation we performed was to measure the difference between the relative phase of the rings at two points separated by an angular region of width 0.78 radians, located symmetrically on either side of the cutting barrier, using the procedure described in section~\ref{phasemeasurement}. Figure \ref{fig:flucvsloadvariance} shows a comparison of the results with a cutting barrier in the outer ring versus the case where both rings were smooth and continuous, as the number of pairs in both rings is varied. The mean of the measured phase difference for both the cases remains close to zero, for all pair numbers. An important observation is that for the case where no cutting barrier is present, the standard deviation of the phase difference across the test region is consistently measured to be much less than $\pi$, even as the number of pairs is reduced to less than 1500. While our phase measurement procedure can become unreliable when the signal to noise ratio in the absorption images falls too low, this consistent result for the connected ring indicates that it is still sufficiently reliable throughout our experimental conditions.

In both cases, the standard deviation of phase difference across the cutting barrier increases as the number of pairs in the system is reduced, with the fluctuations in the cut ring significantly exceeding those of the connected ring. The most interesting feature of Fig.~\ref{fig:flucvsloadvariance} is the dramatic increase in fluctuations when the number of pairs falls below approximately 2500. This increased likelihood of observing a phase difference on the order of $\pi$ points to the onset of the phase-fluctuating regime in the cut ring. The impact of these fluctuations is still evident in the data for the connected ring case, however, because our experimental procedure for creating a connected ring involves first evaporating in a cut (and potentially fluctuating) ring, and reconnecting it before release. We found that the probability for observing a non-zero winding number in a (re)connected outer ring under these same conditions was $0.16^{+0.28}_{-0.01}$ for pair numbers below 2500, and $0.00^{+0.01}_{-0.00}$ above that number. Taken together, these results strongly suggest that phase fluctuations are responsible for difficulties in reliably initializing the system in a selected current state in the low-density limit.

\begin{figure}[!t]
\begin{center}
\includegraphics[width=\columnwidth]{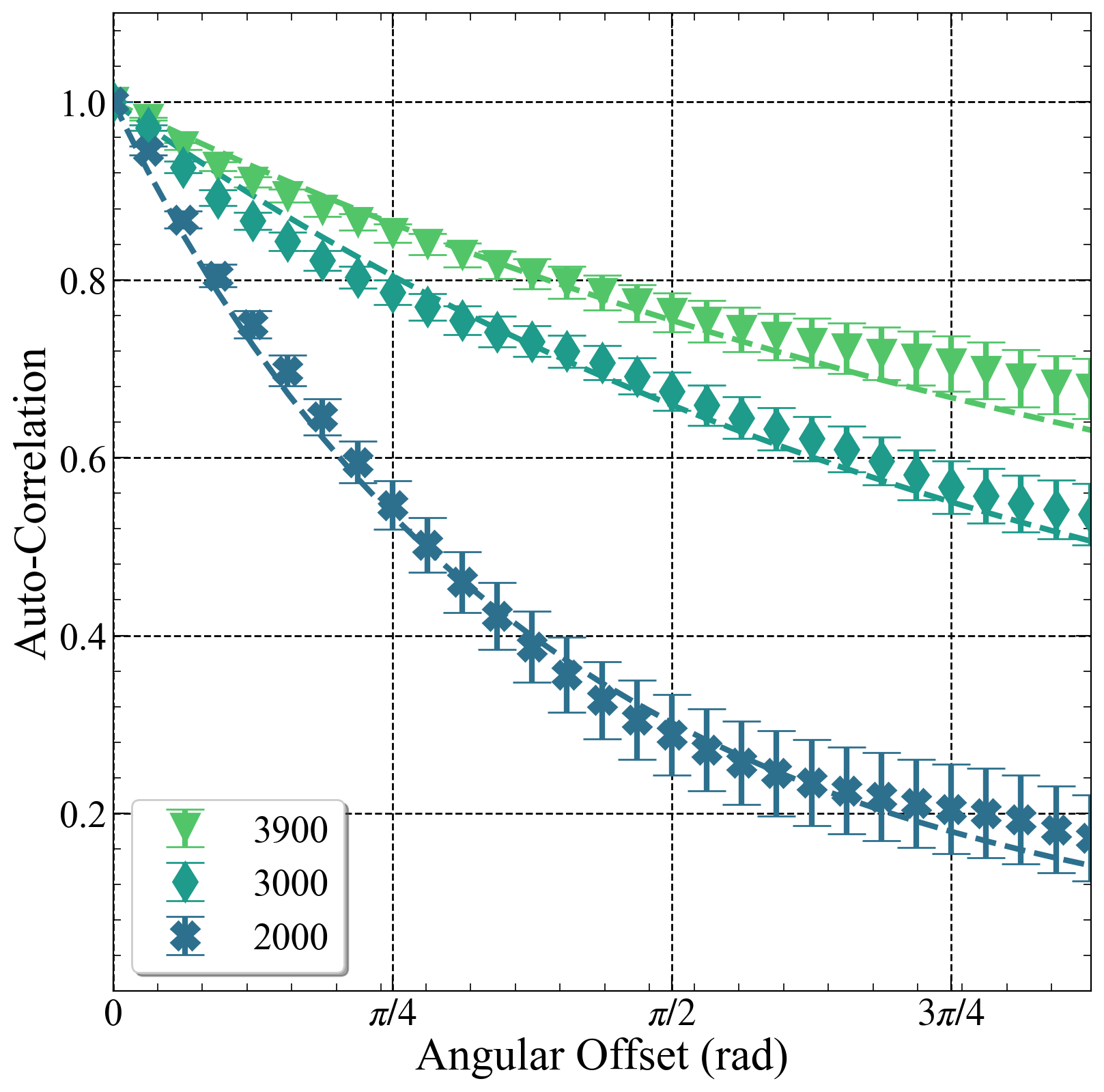}
\caption{Change in the auto-correlation of the phase difference between a connected inner ring and a cut outer ring, as the total number of pairs in both rings is varied. The curves are averaged results for approximately 60 experimental runs each. The average pair number for each bin is indicated in the legend, and the bin width of 1000 pairs is similar to the shot-to-shot number variation for the experiment. Data for another bin centered at 1000 pairs follows the overall trend, but have been omitted for visual clarity. The error bars on each curve are the standard error of the mean. The curves are fit using Eq. \ref{autocorexplicitexpressionfinal2}, and the measured scaled correlation lengths are given in Fig. \ref{fig:correlationlengths}. The x-axis is curtailed at roughly half the domain of the auto-correlation function, where the data is not reliable due to limited overlap between the two copies of the phase profile that are being correlated.}
\label{autocorvsatomno}
\end{center}
\end{figure}

\begin{figure}[!t]
\includegraphics[width=0.45\textwidth]{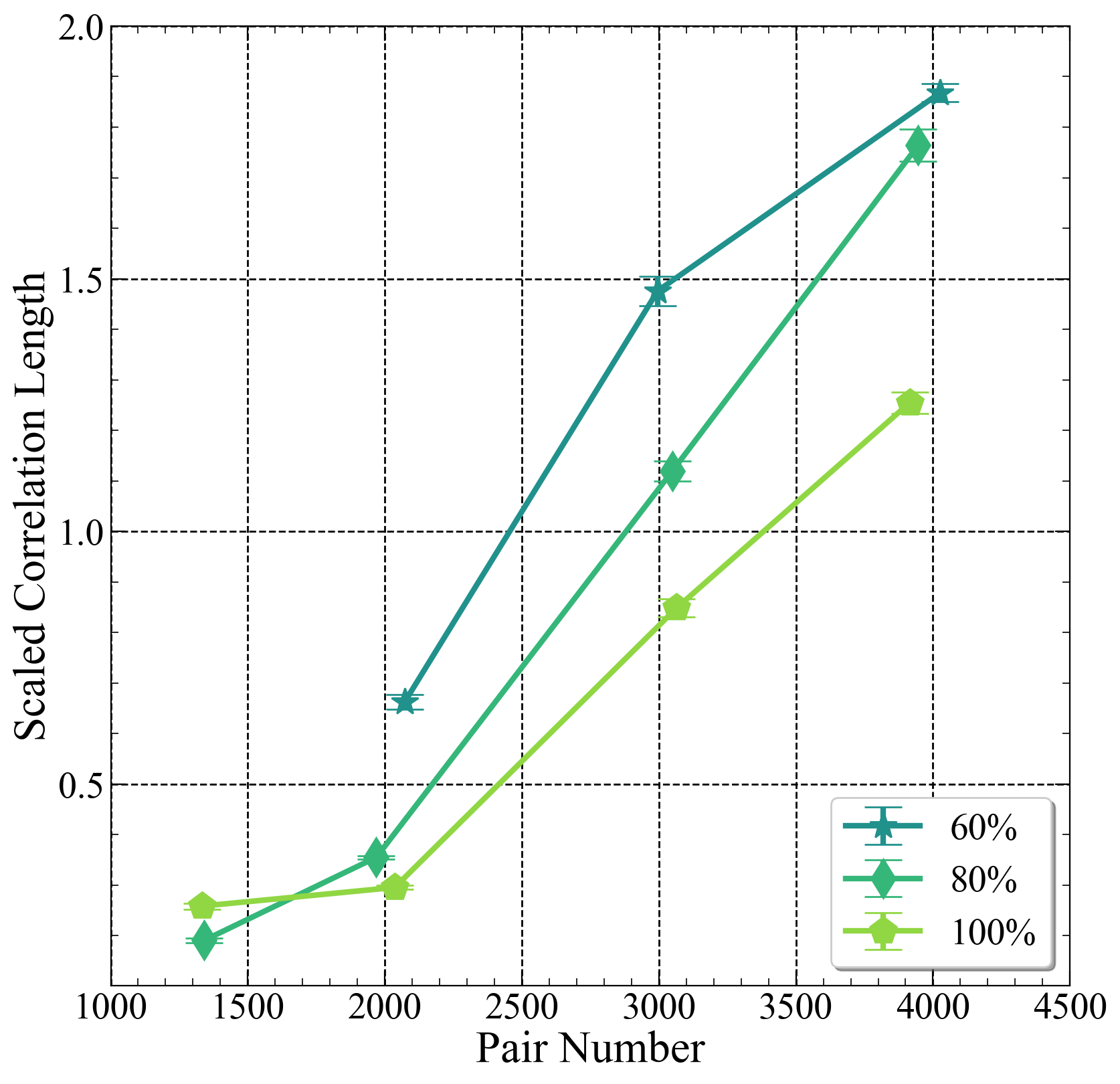}
\caption{Measured scaled correlation lengths $\tilde{l}_{\phi,2}$ for the outer ring, for various barrier strengths and pair numbers. The data for the 90$\%$, 70$\%$ and 50$\%$ cases follow the overall trend, but have been omitted for visual clarity. The error bars are 1-$\sigma$ confidence intervals obtained from the fit covariance matrix.} 

\label{fig:correlationlengths}
\end{figure}

We can obtain a more comprehensive picture of the effect of phase fluctuations at different length scales by calculating the angular auto-correlation of the relative phase of the rings for an individual realization as defined by Eq.~\ref{autocor_equation_linear}, then averaging over many realizations. Fig.~\ref{autocorvsatomno} shows an example set of data obtained for the case of a cut outer ring, as the total number of pairs in the system is varied. As we decrease the pair number, we see a clear reduction in the coherence and the correlation length, as predicted by Eq.~\ref{correlationlenrelation}. We fit these auto-correlations to Eq. \ref{autocorexplicitexpressionfinal2} to extract the scaled correlation length, which we plot in Fig.~\ref{fig:correlationlengths}. A section of the phase profile around the barrier, with a total angular width of 0.78 radians, is excluded in the computation since we do not expect the \textit{in-situ} phase to be well-defined for the cut ring in that region. We see that the measured auto-correlations are well fit by the model over most of the angular domain. At angular offsets on the order of $\pi$ radians, we see deviation from our fit function due to the declining reliability of the auto-correlation owing to the limited overlap between the two copies of the phase profile that are being correlated. Another notable feature of Fig.~\ref{autocorvsatomno} is the ``plateau" observed for the lower pair number cases at offsets larger than $\pi/2$. We suspect this plateau is an effect of coupling both between the rings and between the endpoints of the cut ring, but further investigation will be needed to confirm this conjecture. 

\begin{figure}[!t]
\begin{center}
\includegraphics[width=\columnwidth]{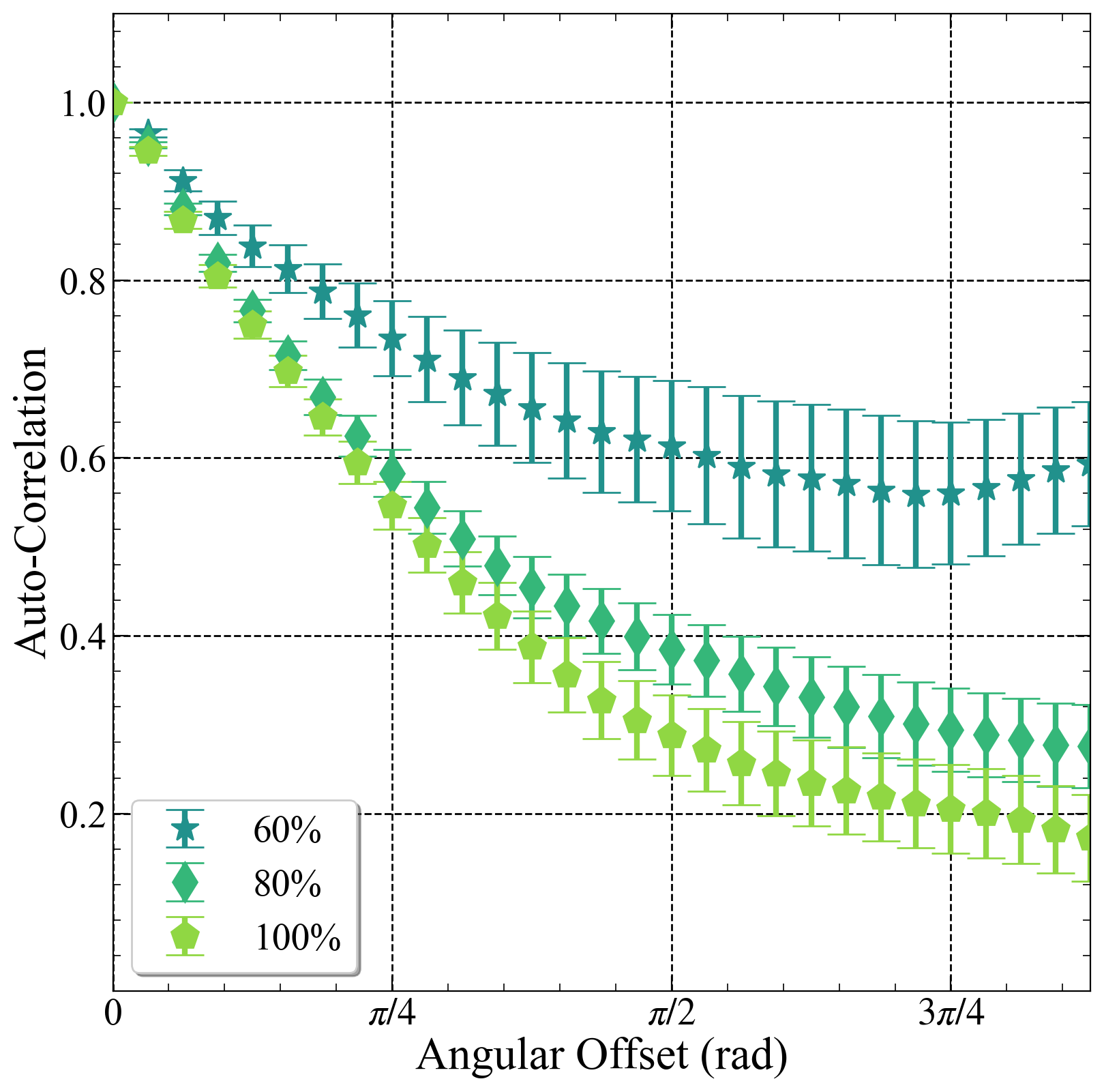}
\caption{Change in auto-correlation of the measured phase difference between a connected inner ring, and a cut outer ring, as the strength of the inter-ring barrier is reduced from its maximal value to the percentage indicated in the legend, for comparable values of total pair number. The figure shows increasing coherence as the inter-ring barrier is successively weakened. The data for the 90$\%$,70$\%$ and 50$\%$ cases follow this overall trend, but have been omitted for visual clarity. The curves are the averages of results from 15,69 and 63 experimental runs respectively, with the total number of pairs for each run falling within a bin width of 1000, centered around 2000. The error bars are the standard error of the mean of the observed statistical variations. The x-axis terminates at roughly half the domain of the auto-correlation function, similar to Fig.~\ref{autocorvsatomno}.}
\label{autocorvsbarrirvsatomno}
\end{center}
\end{figure}

Since the $^6$Li pairs are confined with an attractive red-detuned potential, the inter-ring region can become populated by paired or unpaired atoms if the chemical potential exceeds the ring trap depth. This in turn results in a non-negligible amount of coupling between the two rings. We observe the effect of this inter-ring coupling on the phase fluctuations, by repeating our experiment with the inter-ring barrier weakened by successively larger amounts. Fig.~\ref{autocorvsbarrirvsatomno} shows the auto-correlation where the total number of pairs in the system is held fixed at approximately 2000, but the inter-ring barrier is incrementally weakened by a factor of up to 50\%. This pair number corresponds to the regime where the density in the inter-ring region is effectively zero for the 100$\%$ barrier case, making it ideal for comparing the effects of the inter-ring coupling. Our expectation is supported by the observation of increasing coherence as the inter-ring barrier is weakened. This implies that for higher pair numbers, where inter-ring coupling is more pronounced, even for the 100$\%$ barrier, our correlation length measurement exceeds any prediction based on the assumption of independent rings. It is difficult for us to separate the effects of increasing pair density and inter-ring coupling at present, and a more sophisticated model is required to understand their individual contributions.

The auto-correlations can be fitted to the explicit expression Eq.~\ref{autocorexplicitexpressionfinal2} for interference between a connected and a cut ring, allowing us to measure the correlation length for the outer ring. The scaled correlation lengths for the outer ring for various configurations are plotted against the number of pairs in Fig.~\ref{fig:correlationlengths}. For similar pair numbers, the apparent correlation length is significantly higher for a weaker inter-ring barrier. As the correlation length begins to exceed the circumference of the rings, the precision of our estimation of the correlation length reduces as superfluid coherence extends around the ring and a true condensate is realized.

One aspect of this work that may be of practical significance in the future is the possibility of estimating the temperature of the system from measurements of the phase fluctuations. The model outlined in section~\ref{theorysim} is only a valid description for a single ring, and we would expect it to be most accurate in describing our experiment in the limit of low numbers and maximal height of the inter-ring barrier. Using Eq.~\eqref{temp_equation}, the value of the best-fit temperature was 52(2)~nK, yielding $T/T_c\approx0.9$ given an estimate of the unitary superfluid critical temperature $T_c\equiv0.167T_F=60(5)$~nK~\cite{Ku2012}. We are encouraged by the result still being below the critical temperature, in spite of the fact that noise and other effects would tend to result in a higher temperature estimate.  While we do not presently have a reliable way of independently measuring the temperature to confirm this result, this will be possible in future experiments that combine support for using other thermometry techniques with the capability of controllably increasing the system temperature by a known amount.

\section{Conclusions}
We have performed the first systematic analysis of phase fluctuations in a superfluid under periodic boundary conditions, doing so under conditions where the correlation length is similar to or smaller than the system size. Employing a Fourier analysis-based algorithm, we successfully extract information about the azimuthally varying \textit{in-situ} phase of a superfluid in a ring potential. This was achieved through matter-wave interferometry with a reference ring and analyzing the interference pattern after time-of-flight expansion. Increasing the pair density and inter-ring coupling increased the coherence in the ring, as determined from the auto-correlation of the angular relative phase between the two rings. By measuring the effective coherence length for the outer ring under varying conditions, we observed a transition where the correlation length surpasses the system size. We also found our procedure for initializing a zero-current state in a superfluid ring becomes more successful as the magnitude of phase fluctuations decreases.

Using a simple model for phase fluctuations in independent rings, we obtained a temperature estimate of 52(2)~nK for the system in the limit of minimal inter-ring coupling, which is reasonable in comparison to the estimated critical temperature of 60(5)~nK. A more precise analysis of the temperature dependence of the phase fluctuations could provide a useful means of thermometry for elongated superfluid systems. This work also raises the possibility of investigating the effect of inter-ring coupling, with a potentially unique opportunity to study Josephson vortex dynamics. Alternatively, the influence of inter-ring coupling could be entirely removed by employing a blue-detuned repulsive trap. In future experiments it should be possible to realize even narrower rings with lower atom numbers and reach the quasi-1D limit. In this regime thermal and even quantum phase fluctuations can become significant, and their measurements may provide valuable insight into the behavior of low-dimensional quantum systems.

\section{Acknowledgments}
We thank Roberto Onofrio and Danelle Akanova for helpful discussions and careful reading of the manuscript. This work was supported by the National Science Foundation (Grant No. PHY-2046097).

\bibliography{main}

\begin{thebibliography}{31}
\bibitem [{Fer()}]{FermiEnergyNote}%
  \BibitemOpen
  \href@noop {} {}\bibinfo {note} {{The Fermi energy of a trapped unitary Fermi gas can be estimated as $E_F=\beta^{-1/4}\hbar\omega_z\sqrt{2\pi a_z^2n_2}$, with $\beta\approx0.38$ the Bertsch parameter, $\omega_z$ the axial trap frequency, $a_z^2\equiv\hbar/M\omega_z$ and $n_2$ the peak column density.}}\BibitemShut {Stop}%
\bibitem [{std()}]{stddevnote}%
  \BibitemOpen
  \href@noop {} {}\bibinfo {note} {The standard error of the sample standard deviation approaches 1/$\sqrt{2(n-1)}$ for large $n$, and is a reasonable approximation for $n>10$}\BibitemShut {NoStop}%
\end{thebibliography}

\end{document}